\documentclass[useAMS,usenatbib]{mn2e}
\usepackage{graphicx}
\usepackage{natbib}
\newcommand{\apjs}{ApJS}

\newcommand{\apjl}{ApJL}

\newcommand{\aap}{AAP}

\title[Disc streams, the Galactic potential and its perturbations]{Stellar disc streams as probes of the Galactic potential and satellite impacts}
\author[Laporte, Johnston \& Tzanidakis]{
\parbox[t]{\textwidth}{Chervin F. P. Laporte$^{1,2}$\thanks{$\!\!$Simons Fellow}$\thanks{$\!\!$ CITA National Fellow; e-mail:cfpl@uvic.ca}$, Kathryn V. Johnston$^{1}$, Anastasios Tzanidakis$^{1}$}\\
\\
$^{1}$ Department of Astronomy, Columbia University, 550 West 120th Street, New York, NY, 10027, U.S.A\\
$^{2}$ Department of Physics and Astornomy, University of Victoria, 3800 Finnerty Road, Victoria, B.C., V8P 4HN, Canada\\
}
\begin{document}
\date{}
\pagerange{\pageref{firstpage}--\pageref{lastpage}} \pubyear{2011}
\maketitle
\label{firstpage}
\begin{abstract}

Stars aligned in thin stream-like features (feathers), with widths of $\delta\sim1-10^{\circ}$ and lengths as large as $\Delta l\sim180^{\circ}$, have been observed towards the Anticenter of our Galaxy and their properties mapped in abundances and phase-space.
We study their origin by analysing similar features arising in an N-body simulation of a Galactic disc interacting with a Sagittarius-like dwarf spheroidal galaxy (Sgr). 
By following the orbits of the particles identified as contributing to feathers backwards in time, we trace their excitation to one of Sgr's previous pericentric passages. 
These particles initially span a large range of phase-angles but a tight range of radii, suggesting they provide a probe of populations in distinct annuli in the outer Galactic disc.
The structures are long lived and persist after multiple passages on timescales of $\sim4 \,\rm{Gyrs}$. On the sky, they exhibit oscillatory motion that can be traced with a single orbit mapped over much of their full length and with amplitudes and gradients similar to those observed. We demonstrate how these properties of feathers may be exploited to measure the potential, its flattening, as well as infer the strength of recent potential perturbations.

\end{abstract}
                                                                                                                                                                                                                                                                                     
\section{Introduction}The stellar halo of the Galaxy is regarded as a testament to the hierarchical build up of galaxies expected in the current cosmological paradigm. It is characterised by a smooth component of stellar material from progenitor galaxies accreted long-ago that are now phase-mixed as well as streams and shells from more recent accretion events \citep{Bullock2005, johnston08}. The era of large surveys such as the 2MASS and Sloan Digital Sky Survey opened a new door in characterising the stellar halo leading to the discovery of a dozen more satellites to the Milky Way in addition to the classical known dwarfs \citep{belokurov06a,belokurov07} showing that the stellar halo inventory is made up of streams from ancient shredded dwarf galaxies as well as cold streams of globular clusters \citep{newberg02,majewski04,belokurov06,bell08,bonaca12}. \\
\\
These same surveys also revealed the existence of an extended low-latitude large-scale structure towards the Anticenter, known as the Monoceros Ring \citep{newberg02, ibata03} or ``Galactic Anticenter Stellar Structure" \citep{majewski03}, a vast collection of stars above and below the midplane of the disc between $-30^{\circ}\le b\le30^{\circ}$. The Pan-STARRS survey imaged Monoceros in its entirety, revealing a complex morphology comprised of a large-scale homogeneous overdensity and a succession of sharp overdensities in the form of arc-like features (referred to as ``feathers") which blend smoothly in the disc closer to the Galactic Plane \citep{slater14}. Moreover, the spatial distribution of these arcs is not symmetric about the midplane. Many of these features were already reported in \cite{grillmair06} and \cite{grillmair11}, better known as the Anticenter Stream (ACS) and Eastern Banded Structure (EBS). These fine structures were initially thought to be streams from tidally disrupted dwarfs. However, there is increasing evidence many low-latitude features such A13 \citep[a possible extension of the ACS in red giant branch stars]{sharma10}, or TriAnd overdensities \citep{rocha-pinto04, martin07, sheffield14,deason14} constitute kicked-up disc stars. This is supported from stellar population measurements \citep{price-whelan15, sheffield18}, chemical abundances \citep{bergemann18} and the recent numerical models of the MW-Sgr interactions \citep{laporte18}. Moreover, \cite{deboer17} and \cite{deason18} have recently presented proper motion measurements for the ACS and EBS, respectively showing that these structures have disc-like kinematics similar to Monoceros.\\
\\
On the theoretical side, not much work exists on how these thin stream-like structures can form from the disc. \cite{gomez15b} briefly reported the existence of tentatively similar features in their simulation of a disc disturbed by an infalling satellite\footnote{\cite{slater14} also noted that the simulations from \cite{kazantzidis08} produced similar features to the ACS.}, but did not explore further their origin or what these may look like in phase-space. In this contribution, we seek to determine the physical origins of thin stream-like features as seen around the Monoceros Ring, their stellar population content (in terms of ``birth radii"), understand how they might be used to probe Galactic properties and interaction history, and give predictions for their motion on the sky in light of the upcoming Gaia DR2 mission. This paper is organised as follows. In section 2, we discuss the simulation used for this study. In section 3, we identify ACS/EBS-like structures in observational space and trace their origin and evolution within the disc. In section 4, we show how these can inform us about the potential and discuss their potential to infer past potential fluctuations. In section 5, we present predictions of the expected kinematic properties of feathers in observational space. We discuss our results in light of upcoming astrometric and spectroscopic missions and conclude in sections 6 and 7 respectively.

\section{The simulation}

The simulation used in this study is part of the study from \cite{laporte18}. Briefly, this is a live N-body model following the interaction of a massive Sgr-like dSph with a Milky Way-like Galaxy. The host MW has the following properties: a dark halo of $M_{h}=10^{12}\,\rm{M_{\odot}}$ with a scale radius, a single exponential disc of $M_{disc}=6\times10^{10}\,\rm{M_{\odot}}$, with a scale radius of $R_{d}=5.3\,\rm{kpc}$ and scale height of $h_{d}=0.35 \,\rm{kpc}$ and a central bulge with a mass of $M_{bulge}=10^{10}\,\rm{M_{\odot}}$. The mass model is subject to adiabatic contraction, which steepens the dark matter profile to lead to a final mass model with a circular velocity of $V_{circ}=239 \,\rm{km/s}$ at $R_{0}=8 \,\rm{kpc}$. The simulation ends with a final remnant of $M_{Sgr}\sim3\times10^{9}\,\rm{M_{\odot}}$ with a central velocity dispersion of $\sigma\sim20 \,\rm{km/s}$, producing streams that roughly follow the M-giants of \cite{majewski03}, insuring some reasonable degree of realism.

\begin{figure}
\includegraphics[width=0.5\textwidth,trim=0mm 0mm 0mm 0mm, clip]{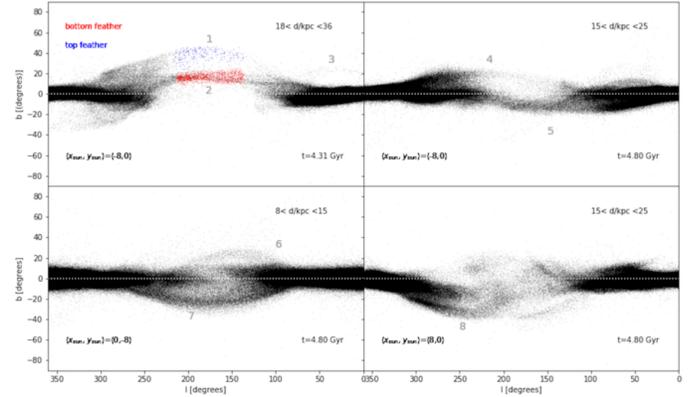}
\caption[]{The Galactic Anticenter from solar neighbourhood-like regions and times in the simulation. Following Sgr's first pericentric passage ($t_{peri}=1.7\,\rm{Gyr}$), the outer disc gets perturbed producing Monoceros-like distribution of stars as well as thin feathers, stream-like structure similar to the ACS, EBS. These are not symmetric about the midplane, have standard deviations of $1^{\circ}-5^{\circ}$ in width, with varying lengths and smoothly blending into the disc. These structures are number from 1 to 8 in the different panels. Two structures are selected for orbital follow-up, marked in blue (top feather) and red (bottom feather) respectively in the top left panel.}
\end{figure}

\section{Disc streams at the disc/halo interface}

\subsection{Identification of feathers}
Our simulations suggest that throughout much of the evolution of the Galaxy, Sgr excites vertical perturbations \citep{laporte18}. Monoceros-like looking features are excited at each passage from the midplane as m=1 arc-like vertical patterns which wind-up gradually. In this orchestration, we also see the rise of stream-like features at the extremeties of the Monoceros-like regions, showing up as thin density enhancements, when viewed as an observer. These are ubiquitous throughout the simulation. In Figure 1, we present a series of snapshots as seen from solar-neighbourhood-like regions in observational space of latitude and longitude $(l,b)$. We note several similarities and resemblances with those features observed in the Pan-STARRS data \citep{slater14} and earlier observations using the SDSS from \citep{grillmair06,grillmair11}. The feathers' widths are {\it small}, spanning $\delta\sim10^{\circ}$ on the sky and they are {\it elongated} covering most of the Anticenter up to $l\sim150^{\circ}-200^{\circ}$. Their shape is distorted with a peak in height and they smoothly reconnect back to the midplane of the disc. The standard deviation measured in the widths of our simulated structures vary from $1^{\circ}$ to $5^{\circ}$ degrees which are consistent with the widths reported by \cite{grillmair06}. In order to better understand the origin of these structures, we will focus on two feathers, lying unambiguously above one another in the top left panel of Figure 1. In the next section we will follow their orbits in order to identify the cause of their dramatic appearance on the sky.

\subsection{Tidal excitation of feathers}

\begin{figure}
\includegraphics[width=0.5\textwidth,trim=0mm 0mm 0mm 0mm, clip]{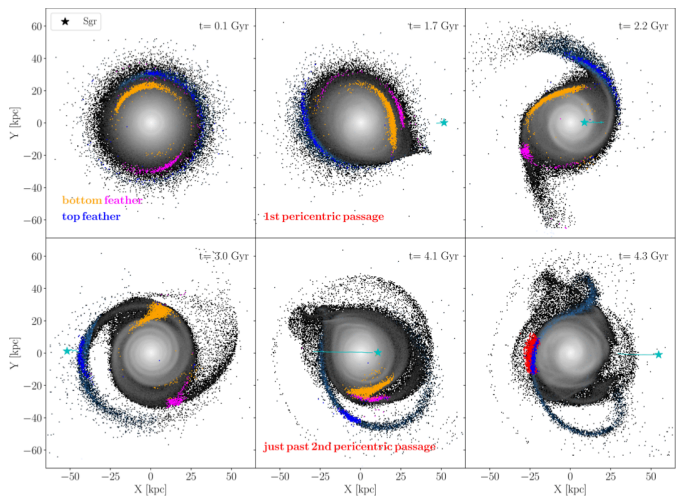}
\caption[]{Evolution of the Galactic disc and particles associated with the top (dark blue points) and bottom (red points) feathers visually identified in Figure 1. We note that the top feather (blue points) is always clearly associated with a single tidal tail, while the bottom feather consists mainly of a superposition of two distinct groups of stars (magenta and orange points) which only meet by $t=4.3\,\rm{Gyr}$ when they were selected (red points). When traced back to their initial positions, the feather stars span a wide range of phase angels, but tight range of radii. As the Sgr-like dwarf makes its first pericentric passage at $t=1.7 \,\rm{Gyr}$, the disc warps and excites the formation of tidal arms. The feather member stars get confined to the tidally excited arms and with their spread in phase angle gradually decreasing. Although, Sgr makes its second pericentric passage at $t=4\,\rm{Gyr}$, the structures that are seen by $t=4.3\, \rm{Gyr}$ are reminescent of an anterior interaction.}
\end{figure}

\begin{figure}
\includegraphics[width=0.5\textwidth,trim=0mm 0mm 0mm 0mm, clip]{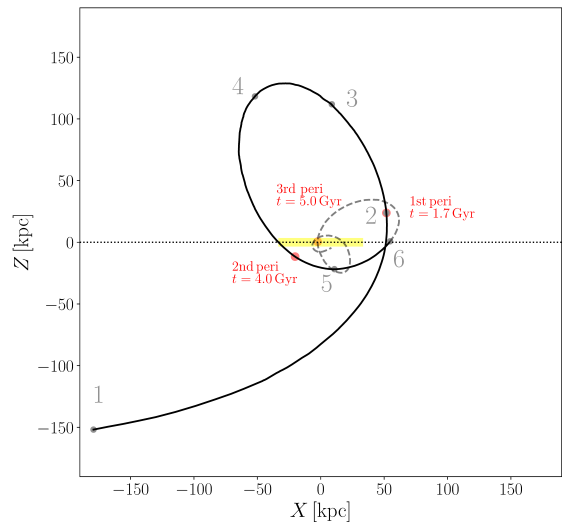}
\caption[]{Orbit of the Sgr-like satellite perturbing the disc prior (solid black line) and after (dashed black line) identification of the feathers at $t=4.3\,\rm{Gyr}$ . The different grey points overlayed on Sgr's orbit corresponding to the positions the satellite is at for each snapshots of Figure 2. The positions and times when Sgr makes its first three pericentric passages are marked in red. For illustration purposes, a thick yellow line of 60 kpc diameter is placed at $(X,Z)=0$ to represent the disc.}
\end{figure}

Having identified several ACS-like features in our simulations of satellite/disc interactions, we now ask what these features correspond to in physical space and specifically what is their origin? To this end, we focus ourselves on two feathers we trace back in time in order to gain insight into their formation before the disc was perturbed by Sgr. Figure 2 shows a series of snapshots of the disc's surface density overplotted with the stars belonging to the feathers identified in Figure 1. We also show the location of Sgr's main body in the $(X,Y)$  plane along with its projected trailing orbit. To further supplement Figure 2, we present the orbit of Sgr in the $(X,Z)$ plane in Figure 3, for which we mark its location for all six panels by grey numbers and its first three pericentric passages in red. 

\begin{figure}
\includegraphics[width=0.5\textwidth,trim=40mm 10mm 40mm 10mm, clip]{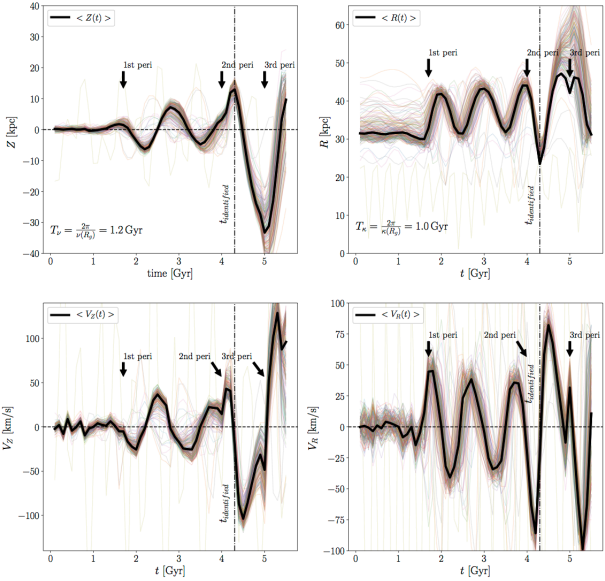}
\caption[]{Motion in Z and R coordinates of the top feather member star particles identified in Figure 1 and 2. {\it Left Panel:} Orbit in Z-component of each feather star (thin coloured solid lines) and of the median population (thick black solid line). The dash-dotted line marks the time at which the stars were observed and selected. The harmonic motion imparted of the stars and sudden changes of amplitude that ensue are correlated with each subsequent pericentric passage. Later pericentric passages do not affect significant change in the dynamics of the feathers. The period of vertical oscillations corresponds to $T=2\pi/\nu=1.2\,\rm{Gyr}$. {\it Right Panel:} Orbit in  cylindrical $R$-component of each feather star (thin coloured solid lines) and of the median population (thick black solid line). The period of radial oscillations corresponds to $T=2\pi/\kappa=1.0\,\rm{Gyr}$. Omitting a few spurious stars, the bulk of the feather stars span a tight range of radii, throughout the whole interaction with the Sgr up to the time they are selected and beyond. This would suggest that one would expect feathers to have a tight spread in abundances. Feathers roughly follow an epicyclic motion.}
\end{figure}

Visual inspection of Figure 2 shows that, initially, prior to Sgr's first pericentric passage, the stars span a wide range of phase angles but tight range of radii. As the first pericentric passage of Sgr proceeds ($r_{peri}=50\,\rm{kpc}$, $t=1.7 \,\rm{Gyr}$), it warps the disc and excites spiral structure and vertical oscillations \citep{purcell11, gomez13, laporte16,laporte18}.  We see that feathers group themselves to become parts of tidal tails from the passage of Sgr about the midplane. This is particularly discernable for the top feather (marked in blue) for which we also tagged in light blue the full tidal tail (identified at $t=3.0\,\rm{Gyr}$) it belongs to. We will see in section 5 that this tail/arm can be fully mapped kinematically. As the simulation continues, a second Sgr pericentric passage occurs ($t=4.0 \,\rm{Gyr}$, $r_{peri}=30\,\rm{kpc}$), this time physically crossing the disc as seen in Figure 3. However, we note that the feathers that are observed at the time of identification were produced during Sgr's earlier (first) pericentric passage. Because of the longer timescales in the outer disc, it is thus possible to presently witness structures that have not yet fully phase-mixed but were excited long ago. We also note that the interpretation of these feathers is not always trivial, even within the simulations. For example, the bottom feather identified in Figure 1 as red stars, turns out to consist of a superposition of two distinct structures in the disc, which separate out in the 5th panel, which we colour coded in orange and magenta in Figure 2. 

Our simulations shows that a disc origin to all the features in the Anticenter of the Galaxy can give rise to a superposition of coherent stream-like features (which all rotate) as a result of an interaction with a single satellite\footnote{In principlal, this does not preclude a multiple satellites' origin.}. Kinematical variations across the various structures are thus expected due to the fact that the Galactic Anticenter is constituted of structures with different guiding radii. This would explain why the ACS, Monoceros and EBS all have qualitatively similar yet distinct kinematical properties. We will present kinematical maps in section 5. 

\begin{figure}
\includegraphics[width=0.5\textwidth,trim=0mm 30mm 0mm 30mm, clip]{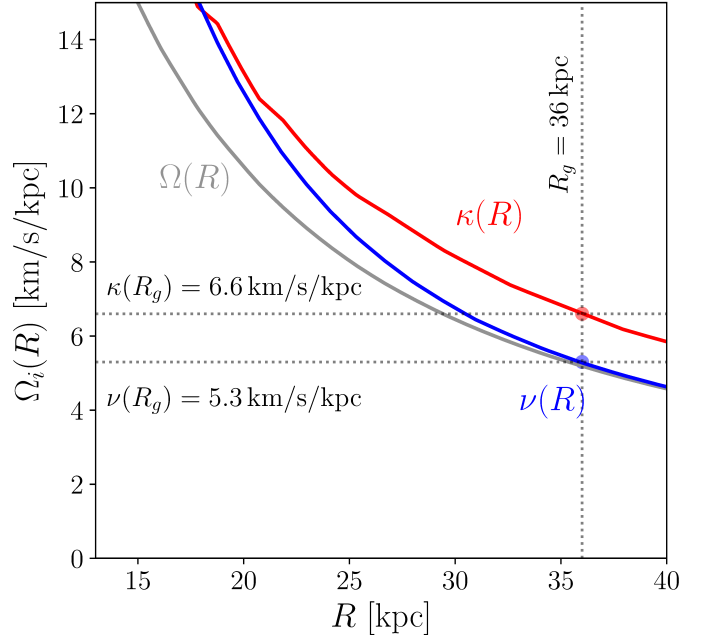}
\caption[]{Angular velocity $\Omega(R)$, epicyclic frequency $\kappa(R)$ and vertical frequency $\nu(R)$ as a function of cylindrical radius $R$ for the given simulated Galaxy mass model, calculated at $t=0.0\,\rm{Gyr}$. The estimated values of $\nu(R_{g})=5.3\,\rm{km/s/kpc}$, $\kappa(R_{g})=6.6 \,\rm{km/s/kpc}$ to describe the equations of motion of the feathers about the guiding radius $R_{g}\approx36\,\rm{kpc}$ through the epicyclic approximation come out exactly as those expected for the given mass model.}
\end{figure}

\subsection{Orbital evolution of feathers and distribution of initial guiding radii}

Deeper insight into the formation, evolution and composition of feathers can be gained by examining the orbits of the stars which constitute it. In Figure 4, we present the motions in the radial and vertical direction of all N-body star particles members ($N_{\rm{members}}\sim2500$) of the top feather identified in earlier figures. 

Before going on to look at the time evolution, one important feature of Figure 4  is the small spread in initial guiding radii of the feather members prior to any interactions. This demonstrates that the structure originates from a narrow annulus of pre-existing disc material and suggests that feathers could 
be used to probe earlier parts of the Galactic disc. 
Interestingly, \cite{bergemann18} presented measurements of abundances of individual M-giants in the A13 and TriAnd and found very narrow spreads in abundances. This indicates that the original annulus in the outer disc that they originated from could have been mono-abundance, rather unlike the more local measurements of the current disc \citep[see e.g.][]{adibekyan12,haywood13,hayden15}. 

Later times in Figure 4 show that the first pericentric passage $t_{peri}=1.7\,\rm{Gyr}$ excites a strong perturbation to the orbits in the vertical direction. As the interaction carries on, a second impulse at $t_{peri}=4.0 \,\rm{Gyr}$ increases the amplitude of the oscillations in the vertical direction. Analogous changes are also seen in the radial direction, for which Sgr imparts a strong change in guiding radius for the stars at first pericenter $t=1.7 \,\rm{Gyr}$ from $R=30\,\rm{kpc}$ to $R=37\,\rm{kpc}$. Interestingly, the excitation of the stars in the feathers is collective and their oscillatory motion is coherent over long time scales $t\sim 4 \,\rm{Gyr}$, showing little scatter about the mean motion of the structure, as shown by the solid black line. 
\footnote{Note that a few stars (less than $1\%$ of all members) do not follow the same motion as all other feather members. This is expected due to our crude observational spatial selection in the $(l,b)$ plane, which will inevitably non-member stars with different histories.} 


\begin{figure}
\includegraphics[width=0.5\textwidth,trim=0mm 0mm 0mm 0mm, clip]{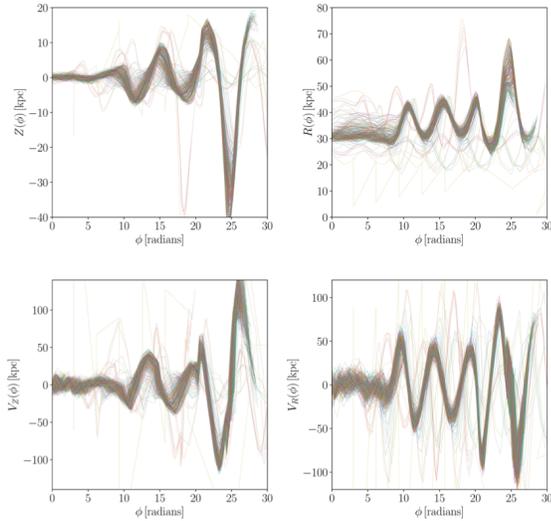}
\caption[]{Evolution of $Z$, $R$, $v_{Z}$ and $v_{R}$ of the top feather member star particles as a function of angle. The motion of feathers is coherent, while the particles reach a certain phase angle at different times they all follow a similar mean pattern.}
\end{figure}

Close inspection reveals that the stars are set on orbits that follow roughly our expectation of vertical and radial oscillations from the epicyclic approximation given by
\begin{equation}
\label{eqn:z}
z(t)\approx Z_{f}\rm{sin}(\nu t +\eta),
\end{equation}
\begin{equation}
\label{eqn:vz}
v_z(t) \approx Z_f \nu \cos(\nu t +\eta)
\end{equation}
and
\begin{equation}
\label{eqn:R}
R(t)\approx X_{f}\rm{sin}(\kappa t +\alpha) + R_{g},
\end{equation}
\begin{equation}
\label{eqn:vR}
 v_R(t)\approx X_{f}\kappa\rm{cos}(\kappa t +\alpha).
\end{equation}
Here $Z_f$ ($X_f$) and $\nu$ ($\kappa$) represent the amplitude and frequencies of the vertical (radial) oscillations about the guiding center at radius $R_g$. $\alpha$ and $\eta$ are constants.
Specifically, the mean motion in $z$ between pericenters 1 and 2 in the top left panel of Figure 4   is well fit for values of $Z_{f}\sim10 \,\rm{kpc}$ and vertical frequency $\nu=5.3 \,\rm{Gyr}^{-1} = 5.3 {\rm km/s/kpc}$. These numbers in turn predict velocity fluctuations up to $\Delta v_{z}\sim Z_{f}\nu=50 \,\rm{km/s}$, as observed in the bottom left panel. Similarly the motion in the radial direction is described by $X_{f}\sim5\,\rm{kpc}$, $\kappa\sim 6.5 \,\rm{Gyr}^{-1} = 6.5 {\rm km/s/kpc}$ about the guiding radius $R_{g}=37\,\rm{kpc}$. From these numbers one would expect radial velocity fluctuations up to $\Delta v_{R}\sim X_{f}\kappa=32.5\,\rm{km/s}$ as seen in the bottom right panels. 

Figure 5 demonstrates that the values estimated for the vertical and epicyclic frequencies, $\nu$, $\kappa$ are in agreement with the values expected for our initial Galaxy mass model at the guiding radius $R_{g}=37\, \rm{kpc}$. 
We stress, that these are estimates and we do not argue that the motion of the stars follow exactly the epicyclic approximation, but can be understood as such to a first order. 

Having shown that the particles in feathers oscillate coherently in time, it is important to understand what their distribution in phase-space can tell us at a single point in time (i.e. in principle for observations).
Figure 6 repeats the quantities plots the same time evolution for each particle in Figure 4 as a function of azimuthal angle in the disc, $\phi$.
The coherent oscillation is still apparent, indicating that feather members share similar azimuthal phases of their turning points. 
We conclude that observations of a single feather can outline the common orbit of its constituent members.
This has very interesting implications for the use of these structures as potential probes.

\section{Feathers as Probes of the Potential and Satellite Impacts}

Traditionally, assessments of the strength of the gravitational  potential in the disc plane have been made using random tracers and requiring that their vertical density distribution be supported by the spread in their vertical motions \citep[e.g.][]{holmberg00}.
These have so far been limited to fairly local regions of the disc where data of sufficient scope and accuracy could be collected to measurement both the density and velocity distribution of appropriate tracers. The prospect of Gaia data offers the possibility of applying this technique more globally \citep[see][for some first results with the current data]{hagen18}.

In contrast, the recognition of feathers as plausibly outlining single orbits (i.e. distinctly {\it non}-random distributions) allows the construction of an alternate potential probe that requires far fewer tracers for analogous accuracy, and no assessment of their density distribution. Moreover, these feathers exist (and our assumptions are most likely to be valid) towards the outskirts of the Galactic disc, precisely where the potential is least well know, and where we might be sensitive to the transition from flattening of the potential due to the baryonic disc and that due to the dark matter halo.

In order to more clearly demonstrate the connection between observable properties of the phase-space locus of a feather (i.e. its position $z_{\rm feather}, R_{\rm feather}$ and velocities $v_{z, \rm feather}, v_{R, {\rm feather}}$ as a function of azimuthal angle $\phi$), its origins and the potential of the Galaxy we consider the event that excited the feather.
Specifically, we assume
velocity changes  $\Delta V_z$ and $\Delta V_R$ were imparted impulsively by a satellite interacting with stars initially spread around a single azimuth (which we label as $\phi_0=0$) on near planar, near-circular orbits a time $T$ ago, and now spread around angular position $\phi_T = \Omega T$. Then we can recast Equations \ref{eqn:z} and \ref{eqn:R} as:
\begin{equation}
\label{eqn:zphi}
z_{\rm feather}(\phi)\approx {\Delta V_z \over \nu} \sin\left({\nu \over \Omega} \phi- \nu T\right)
\end{equation}
\begin{equation}
\label{eqn:vzphi}
 v_z{\rm feather}(\phi)\approx \Delta V_z  \cos\left({\nu \over \Omega} \phi- \nu T\right).
\end{equation}
\begin{equation}
\label{eqn:Rphi}
R_{\rm feather}(\phi)\approx {\Delta V_R \over \kappa} \sin\left({\kappa \over \Omega} \phi- \kappa T\right) 
\end{equation}
\begin{equation}
\label{eqn:vrphi}
v_{R, \rm feather}(\phi) \approx \Delta V_R  \cos\left({\kappa \over \Omega} \phi- \kappa T\right).
\end{equation}
Measures of the amplitude, phase and azimuthal wavelength of a feather in $z$ and $R$ can thus in principle give us measures around the feathers guiding center of the parameters $\Omega$, $\kappa$, $\nu$, $T$, $\Delta V_z$ and $\Delta V_R$. These in turn tell us:
\begin{enumerate}
\item \underline{The flattening of the potential} in the ratio $\nu/\Omega$ (1 for a perfectly spherical potential, $<1$ for an oblate potential and $>1$ for the prolate case); 
\item \underline{The slope of the rotation curve} in the ratio $\kappa/\Omega$ (1 for a Keplerian potential and $\sqrt{2}$ for a perfectly flat rotation curve);
\item \underline{The local angular circular speed}, $\Omega$;
\item \underline{Velocity impulses from the satellite interaction}, $\Delta v_R, \Delta v_Z$;
\item \underline{The time since the interaction}, $T$ .
\end{enumerate}

Of course, reality is more complicated than the picture above as the disc has suffered multiple interactions with Sgr alone. Nevertheless, Figures 4, 5 and 6 already suggest that these simple ideas may be applied to the most prominent features as signatures of the most recent events. 

Figure 7 provides one more simple demonstration of the power in feathers by estimating $\nu$ as a function of $\phi$ along one of our feathers using kinematical information. To do this, we focus on the top feather of Figure 1 which can be distinguished from the background. We offset the average velocities $<v_z>$ measured at a few locations along the feather by 90 degrees (since $\Omega \sim \nu$ in this region of our disc)  and divide by the average heights $<z>$ at those chosen locations. We see that it is in fact possible to recover a nearly constant ratio along $>100^{\circ}$ as an estimate for $\nu\sim5.3\rm{Gyr^{-1}}$. 

\begin{figure}
\includegraphics[width=0.5\textwidth,trim=0mm 50mm 0mm 50mm, clip]{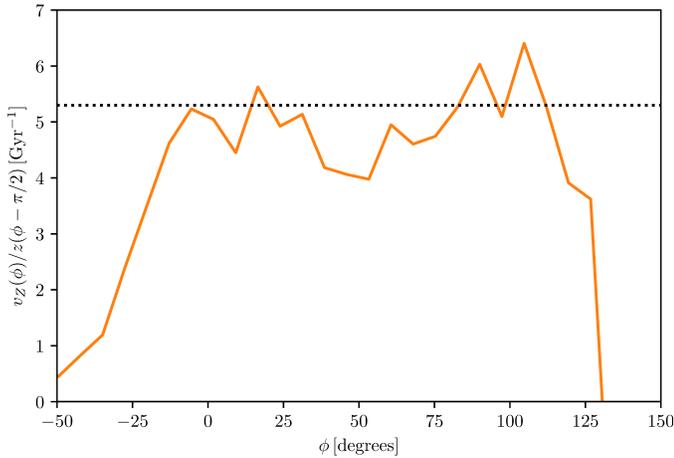}
\caption[]{Ratio of$ Vz/Z,$ for the top feather as a function of $\phi$ along the feather. An almost constant value across $>100^{\circ}$ is recovered over almost the full extent of the feather, nearing the expected $\nu\sim 5.3 \rm{Gyr}^{-1}$.}
\end{figure}

More speculatively, if multiple feathers from a single event (i.e. sharing $T$, but originating from different parts of the Galactic disc) could be identified, the full three-dimensional impulse from the interaction might be reconstructed and the original guiding centers of material now in the feathers might be deduced. A full exploration of these more speculative ideas is deferred for future work. We restrict our focus in this paper to next asking whether current and near future surveys are in a position to provide useful phase-space maps of these structures.

\section{Feather kinematics in observational space}

\begin{figure}
\includegraphics[width=0.5\textwidth,trim=0mm 50mm 0mm 50mm, clip]{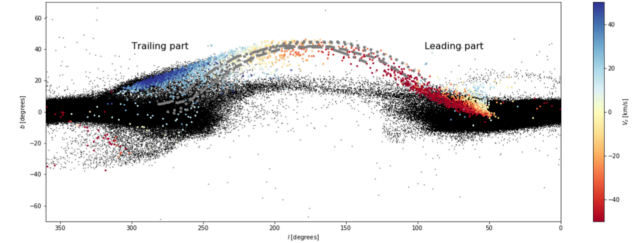}
\caption[]{Past orbit of some of the leading member stars of a feather, for which we colour coded the member stars by their individual vertical velocities. The orbit roughly follows the extent of the feather on the sky}
\end{figure}

Figure 8 re-iterates the result of the previous section, this time in observable co-ordinates, by plotting the past orbits of particles in the simulations projected onto the plane of the sky viewed from an assumed solar position. These projected orbits track the observed positions backwards and forwards along most of the feather. We checked this behaviour as well for a few other feathers.

We now turn to look at the kinematical properties of feathers in the observational space in light of the upcoming Gaia second data release and future complementary ground based surveys which will provide a full 6-D phase-space characterisation of the Anticenter. 
Most importantly, we want to be sure that the feather features we need to measure ($\Delta V_z, \Delta V_R$ and the azimuthal wavelength and phase in each dimension) will be resolved in these surveys.
This would inform us whether there is any potential to quantify the torques that must have acted on the Galaxy in its past. It should also provide us an idea of the kinds of motions expected across the full extent of the ACS and other structures identified in the Pan-STARRS field (Slater et al. 2014) as well a model to compare to recent estimates on portions of the ACS through the Gaia-SDSS derived proper motions of (de Boer et al. 2017; Deason et al. 2018).

\begin{figure*}
\includegraphics[width=1.0\textwidth,trim=20mm 50mm 20mm 50mm, clip]{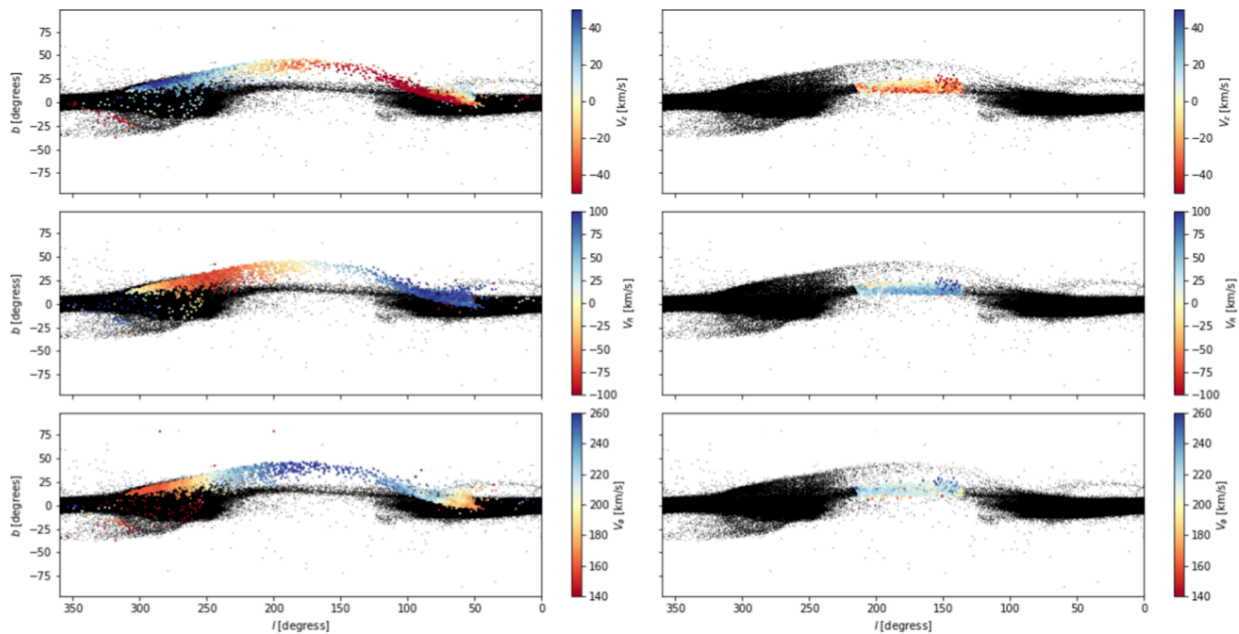}
\caption[]{Vertical, radial and azimuthal motions for 2 identified feathers (left and right panels). The background stars within the same heliocentric distance as the feather members are shown in black. {\it Left panels:} Top feather identified in Figure 1 and 2, showing oscillatory motion across its full length. The changes across the feather are approximately equal to the changes in amplitude in the mean orbit of the feather past second pericentric passage with $\Delta V_{z}\sim 50\,\rm{km/s}$, $\Delta V_{R}\sim75\,\rm{km/s}$. Velocity gradients can also be appreciated at a fixed longitude. {\it Right panels:} Vertical, radial and azimuthal motions for the bottom feather. The lack of a clear oscillatory motion is due to the fact that it consists of a superposition of two separate populations as noted in Figure 2.}
\end{figure*}

Given the simple orbital motion of feathers, and large coverage on the sky, these oscillations may be potentially fully mapped. We now ask how feathers would look like on the sky in velocity space and illustrate the scales of motions which may be measured in the near future with the {\it Gaia} satellite upcoming data release and future ground based complementary surveys (WEAVE, 4MOST, DESI). In Figure 9, we present the vertical, radial and azimuthal motions (top, middle and bottom panels respectively) of the top and bottom feathers (left and right panels respectively) identified in the bottom right panel of Figure 1. For the top feather, for which we have mapped its full structure (light blue points) in Figure 2, a clear sinusoidal signal in the vertical velocities is visible \footnote{We have checked this was the case with other feathers identified in the simulation as well.}, with the apocenter almost coinciding with the maximum peak in latitude (however, in general this needs not be the case due to projection effects). Interestingly, the feathers also exhibit clear gradients in vertical velocity as a function of longitude on scales of $\Delta l \sim 10^{\circ}$ with amplitudes $\Delta v_{Z}\sim 20 \,\rm{km/s}$. Recently, \cite{deboer17} and \citep{deason18} have presented measurements of proper motions for portions of the ACS and EBS streams derived through a combination of SDSS photometry with its 10 year baseline with that of Gaia DR1 with similar values to what we find in the simulations. In particular, \cite{deboer17} finds an oscillating signal as a function of longitude over an extent of $\Delta l\sim100^{\circ}$ with variations in the range $-60 <v_{Z}/\rm{km/s}< 20$. They also measure gradients in latitude the velocity on a scale of $\Delta l\sim10^{\circ}$ with variations of order $\sim 20\, \rm{km/s}$  These numbers qualitatively agree with what we find here and in \citep{laporte18}.

Similarly looking at the radial velocities variations across the feather we can also see velocity gradients which can range up to $\Delta v_{R}\sim25 \,\rm{km/s}$ and clear circular motion. The azimuthal velocities vary between $160\, \rm{km/s}$ and $240 \,\rm{km/s}$, with the peak being reached around the apocenter of the feather. The right panels of Figure 9, shows the bottom feather which also exhibits gradients but are more difficult to discern because of the projection of two separate regions of the disc. Thus while we expect to be able to disentangle feathers in kinematical space, this may not always be possible, especially given the fact that the Gaia satellite will not be able to produce reliable parallaxes at the distances of the Monoceros Ring. However, given that \cite{deboer17} already measure a signal qualitatively and quantitatively close to one of our simulated feathers, it may be possible to fully map feathers in a the near future.

We conclude that the scales of oscillations seen in our simulated data are entirely consistent with those already probed with current data by averaging over many stars.
Looking to the (near) future, {\it Gaia} will provide estimates for the proper motions of individual main sequence turn-off stars (MSTOs) in Monoceros, ACS/A13, of sufficient accuracy which will allow us to assess 6-D phase-space motions with few-ten km/s accuracies and truly map the oscillations in some detail (see appendix A). It should also be noted that tracers such as M-giant stars ($M_{G}\sim-2.5$) and the red clump $M_{G}\sim0.44$ \citep{hawkins17} will be able to trace structures out to TriAnd and measure transverse velocities at the level of $\mathcal{O}(0.1 \,\rm{km/s})$ and $\mathcal{O}(\,\rm {km/s})$ respectively, but also provide estimates of distances in regions where Gaia parallax errors are large, radial velocities and chemical information.

\section{Discussion}

If the Sgr dwarf spheroidal is responsible for the wobble of the MW disc and its diverse identified corrugations as supported by the recent models of \citep{laporte18}, we have identified a new way to explore and interpret properties of tidal arms of the Milky Way (or at least traces of spiral arms/tidal tails of the Galaxy). These should then be visible as thin-streams above the midplane, exhibit kinematics reminescent of the disc and tight spreads in abundances (as a consequence of the tight range of initial guiding radii). There are already indications that this is the case \citep{bergemann18}. In this study the authors present detailed chemical abundances for A13 and TriAnd which had spread in chemical abundances of $0.15\,\rm{dex}$. Whether TriAnd itself might be a feather is difficult to acertain, however the case for A13 seems strong and plausible. A13 structure was identified in M-giants in 2MASS by \citep{sharma10}. The spectroscopic distances derived to A13 place it at a distance of which is similar to the distance estimate to Monoceros and the ACS. A closer look at the location of the M-giants with the MSTOs from the stacked PS-1 image of \citep{bernard16} or that of \citep{slater14} shows that the 2MASS giants at similar distances to the PS-1 MSTOs do trace the ACS and its edge (Appendix B). Thus, the formation of feathers from pre-existing disc material (originally confined to within a tight annulus) through tidal encounters as the ones identified in our simulations is particularly interesting given the tight spreads in abundances and explain in part the result from \cite{bergemann18}. A full confirmation will need to await for a direct connection between the ACS and A13. Complementary spectroscopic surveys to the {\it Gaia} mission such as WEAVE, 4MOST will be particularly relevant to falsify our ``model". 

Because the kinematics and geometry of the feathers is simple, it may be possible using a the combination of kinematics and photometry to use them as tracers of the MW potential and to infer the torques responsible for their motion as a indirect inference method on the mass of the Sgr progenitor. We have shown that this is possible in principle, which is promising given the imminent {\it Gaia} data release. The upcoming release of Gaia DR 2 will reveal proper motions for the entirety of the Monoceros Ring and in doing so will also be able to reveal the full kinematics of feathers such as the ACS and EBS across their full extent, thus opening new possibilities to probe the Galactic potential as well as its evolution.

Not all feathers show clear sinusoidal motions in velocities as these may also be constituted of a superposition of different groups of stars in phase. This opens the possibility that some structures identified in the PanSTARRS data may originate from different separate guiding radii coinciding today at a heliocentric distance of $d\sim10\,\rm{kpc}$ towards the Anticenter. This makes feathers perfect candidates to study their chemical composition, measure chemical gradients in the thin disc populations. Our models are potentially falsifiable which should motivate upcoming spectroscopic surveys to target the ACS, EBS and the Monoceros Ring in the near-future through complimentary surveys to Gaia such as DESI, WEAVE and 4MOST. Spectroscopic surveys will be particularly useful to also disentangle in-situ and accreted components at the halo-disc interface which cannot always be distinguished solely based on kinematics \citep{jean-baptiste17,bonaca17}, thus potentially open new doors to identifying other feathers across the Milky Way.


The appearance of the stream above the midplane in the sky is mainly due to the fact that the structure is seen at apocenter. This is also corroborated by the measurements of \cite{deboer17} who also measure oscillatory motion with velocity gradients in $(l,b)$ with velocities in the ranges $-50\leq V_{z}/\rm{km/s}\leq 50 $. We also show that the orbits of stars in the leading part of the feather trace the full extent of the feather on the sky. Thus given the expected simple motion of feathers, these may be used to not only infer the Galactic potential but also infer the torques on the disc which were inflicted by the Sgr dwarf spheroidal, opening a new window into probing the mass loss history of Sgr. We plan to report on this in more detail and the limitations and possible caveats of feathers in the inferences about the Galactic potential in a forthcoming contribution.

\section{Conclusion}

In this paper we used a live N-body simulation of the interaction of the Milky Way with a Sgr-like dwarf galaxy to study the origin of thin stream-like features, towards the Galactic Anticenter. Our main conclusions on "feathers" can be categorised in three bullet points.

\begin{enumerate}
\item {\bf Nature of "streams" in the Anticenter:} Feathers are thin stream-like which arch above and below the midplane of the disc which smoothly blend into the disc of the MW. They have a narrow width with a dispersion of $\sim1^{\circ}-5^{\circ}$. Physically, they are remnants of tidally excited arms which are excited by previous pericentric passages of a perturbing galaxy, in the case of the Milky Way, potentially the Sgr dSph. These move coherently in phase and can persist for long orbital timescales $\sim 4 \,\rm{Gyr}$ past future pericenters. 

\item{\bf Potential (fluctuations) measurements:} Feathers exhibit simple epicyclic motion. Together with their large extent on the sky, they can be used to directly measure the vertical and potentially epicyclic frequencies of the Galactic potential through the ratios $\nu=\delta v_{Z}/Z_{f}$, $\kappa=\delta v_{R}/X_{f}$. Moreover, the vertical frequency could also be inferred along the stream through the ratio of the vertical velocities and vertical heights which should be roughly constant over several 10s of degrees (in a spherical ideal case this should amount to 180 degrees). Due to the large distances of feathers, the disc self-gravity will be negligible. A value of $\nu(R)=\Omega(R)$ would signal a spherical potential but any deviations $\nu\neq\Omega$, would indicate a flattening of the potential, either instrinsically due to the shape of the MW's dark halo or a more exotic origin such as a dark matter disc. Moreover, the value of $X_{f}$ and $Z_{f}$ and instrinsic disc origin of feathers would allow us to infer the past perturbation to the MW potential responsible for the present motion of feathers. This would turn feathers as more than just a simple tracer of the MW potential but make them actual {\it gravitational detectors} of the MW's past potential fluctuations. Their motion could place strong constraints on Sgr's orbital mass loss history.

\item{\bf Abundance reconstruction of the disc:} Feather member stars span a narrow range of initial guiding radii, suggesting that one should find small abundance spreads in them. This is particularly intriguing in light of the recent measurements of \cite{bergemann18} in A13 and TriAnd. The Anticenter region of the Milky Way towards the region of Monoceros ($d_{helio}\sim10\,\rm{kpc}$) and beyond is populated by a series of thin stream-like features (EBS, ACS, possibly TriAnd). A spectroscopic campaign targetting such features dedicated to measuring abundances across them, combined with a potential inference as well as it fluctuation would open up a new way to reconstruct the initial chemical abundance pattern of the disc. Such a measurement of abundances and inference of initial chemical gradients would put strong constraints on the formation of the thin disc.

\end{enumerate}

\section*{Acknowledgments}
We thank Volker Springel for giving us access to the {\sc gadget-3} code. 
Estimates for {\it Gaia} accuracies were made using the {\tt PyGaia} Python package written by Anthony Brown and we thank him for his work and the Gaia Project Scientist Support Team and the Gaia Data Processing and Analysis Consortium (DPAC) for their support overall. This work made use of {\tt numpy, scipy} and {\tt matplotlib} \citep{numpy, scipy, hunter07}. CL acknowledges support by a Junior Fellow of the Simons Society of Fellows award from the Simons Foundation, a visiting fellow position at the MPA and a CITA National Fellow award. This work used the Extreme Science and Engineering Discovery Environment (XSEDE), which is supported by National Science Foundation grant number OCI-1053575. 
KVJ's work on this project was made possible by NSF grant AST-1614743. CL acknowledges useful discussions with Julio Navarro, Rodrigo Ibata, Facundo G\'omez, Adrian Price-Whelan, Maria Bergemann, Misha Haywood, Paola di Matteo and thanks Eiichiro Komatsu, Simon White for hosting him at the MPA between October and November 2017. CL dedicates this paper to Stephen Hawking for guiding him to pursue a career in astrophysics since 2001.

\bibliographystyle{mn2e}
\bibliography{master2.bib}{}

\appendix

\section{Gaia proper motion errors towards the Anticenter}
In this appendix we assess the role of three different tracers (MSTOs, M-giants and the RCs) in measuring the proper motions of various known structures towards the Anticenter, in particular the AntiCenter Stream / A13 ($d\sim10\,\rm{kpc}$) and TriAnd ($d\sim20-30\,\rm{kpc}$).

\begin{enumerate}
\item {\bf MSTOs:}  We estimate the average MSTOs proper motion errors for Gaia DR2 and the Gaia end-of-misison using a similar methodology to \citep{debruijne14}. MSTOs toward Monoceros ($d\sim10\,\rm{kpc}$) span a magnitude range of $19\leq g\leq 20$ in the SDSS $g$-band and have colours of $g\sim0.3$. Using the colour-colour polynomial fit coefficients in Table 5 of \cite{jordi10} assuming no extinction, these would correspond to magnitudes in the $G$-band in the range $18.6<G<19.6$. At those magnitudes, the expected proper motion errors are in the range $108-255 \,\mu \rm{as/yr}$ which would correspond to velocity errors of $5-12 \,\rm{km/s}$ for individual stars. These would correspond to about $10\%$ velocity errors. Averaging the measurements of individual MSTOs in one $\rm{deg^2}$, which would contain between 100 and 25 stars in the ACS \citep[see Figure 4 of ]{deboer17}, this would result in measurements errors of $1-2\,\rm{km/s}$ to $0.5-1.2\,\rm{km/s}$ which would be at the $1\%$ level.

In the nearer future, with Gaia DR2, one will expect uncertainties in proper motions at $G=17$ of $0.2  \,\rm{mas/yr}$ and at $G=20$ of $1.2  \,\rm{mas/yr}$. Given the magnitude range of MSTOs in the Anticenter, we would expect uncertainties in velocities of about $30-40 \,\rm{km/s}$. Thus through an averaging scheme similar to \citep{deboer17} on a $2 \,\rm{deg^{2}}$ scale, one would recover velocity errors of $2-4 \,\rm{km/s}$ which should be enough to map feathers in quite some detail.

\item {\bf Red clump stars:} Red clump stars are well known standard candles which can provide accurate distances, with an absolute magnitude of $M_{G}\sim0.44$ \citep{hawkins17}. At a distance of $d\sim10\,\rm{kpc}$, these would have an apparent magnitude of $G\sim15.4$. Their expected Gaia end-of-mission proper motion errors on individual RCs would be of about $\sim18\mu\rm{as/yr}$, corresponding to velocity errors $0.85\,\rm{km/s}$. These numbers would be at the percent level from the currently measured signal in stacked MSTOs \citep{deboer17}. Given that Gaia DR2 is expected to provide proper motion errors of order $0.2\, \rm{mas/yr}$ at $G=17$ and $0.06 \,\rm{mas/yr}$ for sources with $G<15$, this means that Gaia DR2 would already provide individual proper motions of order $\mathcal{O}(5 \,\rm{km/s})$. 
\newpage
\item {\bf M-giants:} 
M-giants would sit two magnitude lower to RCs (i.e. $M_{G}\sim-2.5$) and have been mapped towards the Anticenter tracing both the Monoceros/ACS/A13 and TriAnd. At $10\,\rm{kpc}$ and $30 \,\rm{kpc}$, these would have apparent magnitudes between $G\sim12.5$ and $G\sim14.9$. Thus already within the Gaia DR2 release individual M-giants will have proper motions with velocity errors of $3\,\rm{km/s}$. By the end of the mission, we estimate that these individual proper motion errors will be of the order of $0.2-0.9 \,\rm{km/s}$ between Monoceros and TriAnd, thus giving accuracies at the percent/sub-percent level. Complemented with spectroscopic follow-ups it should be possible to have a detailed picture on the structure of the Galactic Anticenter and its feathers.
\end{enumerate}

Thus we concluded that combining all three tracers we should be able in the not so distant future to fully map the Galactic Anticenter in phase-space as well as the space of abundances (and potentially spectroscopic ages with MSTOs).

\section{The ACS in MSTOs and 2MASS M-giants }
In this section we present the full PS-1 data from \cite{bernard16} with the 2MASS M-giants of \cite{majewski04} at situated at the same photometrical distances as the PS-1 MSTOs.  Interestingly the M-giants which were used to characterise A13 trace the ACS structure as well as the abrupt fall-off in density seen in MSTOs above $l\sim40^{\circ}$. It would not be surprising that the two tracers belong to the same common structure, something Gaia DR2 will be able to confirm. A full characterisation of the feathers will require a complementary spectroscopic campaign to complement the stacked MSTO proper motions (and individual M-giants, RCs proper motions) with M-giants radial velocities in order to provide a full 6D view of feathers towards the Anticenter (as these should also provide spectroscopic distances to members, but also abundances).

\begin{figure}
\includegraphics[width=0.5\textwidth,trim=0mm 0mm 0mm 0mm, clip]{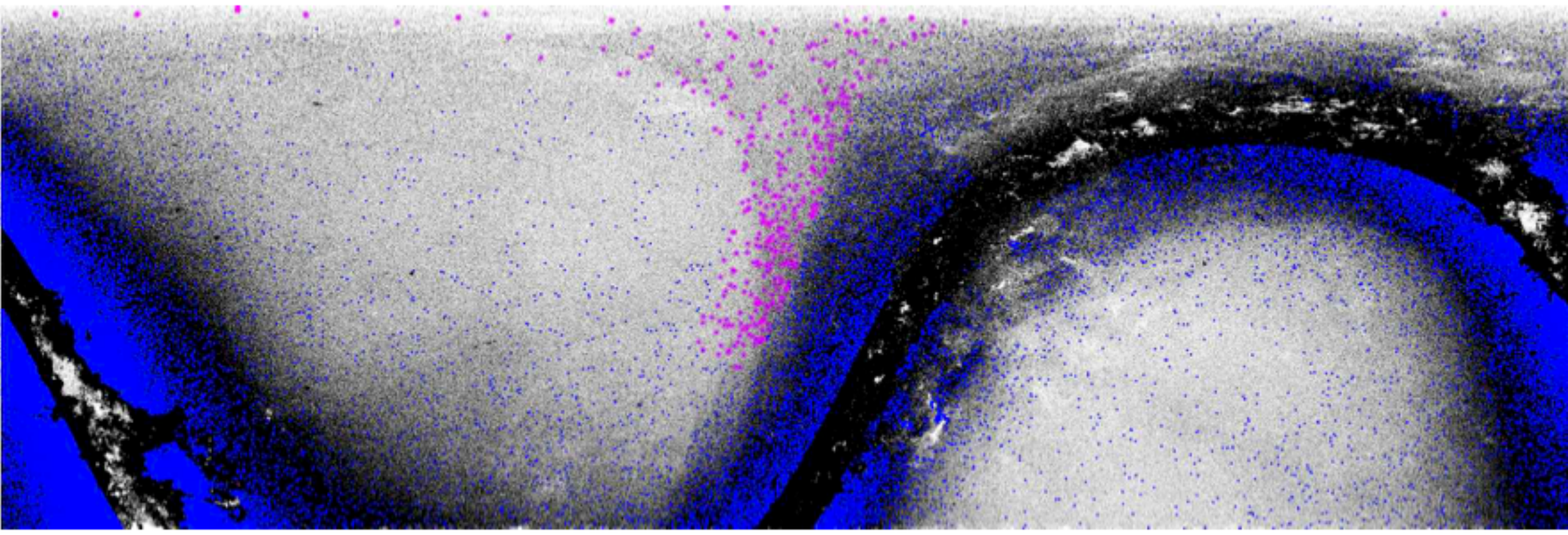}
\caption[]{PS-1 map of MSTOs from \cite{bernard16} overlayed with 2MASS M-Giants at similar photometric distances (blue points). The giants roughly delineating the ACS are shown in magenta.}

\end{figure}

\label{lastpage}
\end{document}